\documentclass[12pt,preprint]{aastex}
\usepackage{emulateapj5}

\usepackage{epsfig}
                                                                                
\def\vkm{km s$^{-1}$}

\def\degree{$^\circ$}

\def\arcs#1{$#1''$}
\def\arcsa#1#2{$#1^{\prime\prime}_{^\textrm{.}}#2$}

\def\solarmass{$M_\odot$}

\def\mJyb{mJy beam$^{-1}$}

\def\Jybk{Jy beam$^{-1}$ km s$^{-1}$}

\def\tlabel#1{(\textit{#1})}

\def\micron{$\mu$m}
                                                                                
\def\ra#1#2#3#4{#1^\mathrm{h} #2^\mathrm{m} #3^\mathrm{s}_{^\textrm{.}} #4}
\def\dec#1#2#3#4{#1\degr #2\arcmin #3^{\prime\prime}_{^\textrm{.}}#4}

\def\VLSR{V_\textrm{\scriptsize LSR}}
\def\Vsys{V_\textrm{\scriptsize sys}}
\def\Voff{V_\textrm{\scriptsize off}}

\def\H2{H$_2$}
\def\N2HP{N$_2$H$^+$}

\def\cCO{C$^{18}$O}

\def\aCO{$^{12}$CO}
\def\bCO{$^{13}$CO}
\def\NH3{NH$_3$}

\def\SOta{$N_J=5_6-4_5$}

\def\putfig#1#2#3{\epsfig{scale=#1,angle=#2,figure=#3}}
\def\putfig#1#2#3{\includegraphics[angle=#1,scale=#2]{#3}}
\def\putfig#1#2#3{}
\def\leftblank#1{}

\begin{document}

\title{Angular momentum loss in the envelope-disk transition region of HH
111 protostellar system: evidence for magnetic braking?}


\author{Chin-Fei Lee\altaffilmark{1,2}, Hsiang-Chih Hwang\altaffilmark{1,2},
and Zhi-Yun Li\altaffilmark{3}} 

\altaffiltext{1}{Academia
Sinica Institute of Astronomy and Astrophysics, P.O.  Box 23-141, Taipei
106, Taiwan; cflee@asiaa.sinica.edu.tw}

\altaffiltext{2}{Graduate Institute of Astronomy and Astrophysics, National Taiwan
University, No.  1, Sec.  4, Roosevelt Road, Taipei 10617, Taiwan}

\altaffiltext{3}{Astronomy Department, University of Virginia, Charlottesville, VA,
USA}

\begin{abstract} HH 111 is a Class I protostellar system at a distance of
$\sim$ 400 pc, with the central source VLA 1 associated with a rotating disk
deeply embedded in a flattened envelope.  Here we present the observations
of this system at $\sim$ \arcsa{0}{6} (240 AU) resolution in \cCO{} (J=2-1)
and 230 GHz continuum obtained with Atacama Large Millimeter/Submillimeter
Array, and in SO (\SOta) obtained with Submillimeter Array.  The
observations show for the first time how a Keplerian rotating disk can be
formed inside a flattened envelope.  The flattened envelope is detected in
\cCO{}, extending out to $\gtrsim$ 2400 AU from the VLA 1 source.  It has a
differential rotation, with the outer part ($\gtrsim$ 2000 AU) better
described by a rotation that has constant specific angular momentum and the
innermost part ($\lesssim$ 160 AU) by a Keplerian rotation.  The
rotationally supported disk is therefore relatively compact in this system,
which is consistent with the dust continuum observations.  Most
interestingly,  if the flow is in steady state, there is a substantial drop
in specific angular momentum in the envelope-disk transition region from
2000 AU to 160 AU, by a factor of $\sim 3$. Such a decrease is not expected
outside a disk formed from simple hydrodynamic core collapse, but can happen
naturally if the core is significantly magnetized, because magnetic fields
can be trapped in the transition region outside the disk by the ram pressure
of the protostellar accretion flow, which can lead to efficient magnetic
braking.  In addition, SO shock emission is detected around the outer radius
of the disk and could trace an accretion shock around the disk. 
\end{abstract}

\keywords{circumstellar matter -- stars: formation --- ISM: individual (HH 111)}

\section{Introduction}

Keplerian rotating disks (KRDs), which are rotationally supported, have been
detected around young stellar objects, especially in Class II and Class I
phases.  Recently, more and more such disks have been found as early as in
Class 0 phase \citep{Lee2009,Lee2014,Murillo2013,Ohashi2014}.  However, the
formation mechanism of such disks is still not well understood because of a
lack of observations at high angular and velocity resolutions.

A few years ago with the Submillimeter Array (SMA) observations, the
rotation profile in the Class I source HH 111 was found to transition from
that conserves angular momentum in the envelope to that of Keplerian in the
KRD \citep{Lee2010}.  This transition was found for the first time in star
formation, providing a clue to the formation mechanism of KRDs.  With
Atacama Large Millimeter/Submillimeter Array (ALMA), recent searches also
found such a transition even in Class 0 sources, e.g., L 1527 IRS
\citep{Sakai2014,Ohashi2014} and HH 212 \citep{Lee2014}.  In order to study
how a KRD can be formed, we need to resolve the transition region, and
compare its structure and kinematics to current model predictions.

Theoretically, in models of non-magnetized core collapse, a KRD can indeed
form as early as in the Class 0 phase \citep{Terebey1984}.  However, a
realistic model should include magnetic field, because recent survey toward
a few Class 0 sources shows that molecular cores are magnetized and likely
to have an hourglass B-field morphology \citep{Chapman2013}.  Unfortunately,
in many current models of magnetized core collapse, the magnetic field
produces an efficient magnetic braking that removes the angular momentum and
thus prevents a KRD from forming at the center
\citep{Allen2003,Mellon2008,Galli2009}.  In those cases, only a flattened
envelope called the pseudodisk can be formed around the central source
\cite[e.g.,][]{Allen2003}.  Magnetic-field-rotation misalignment is
sometimes able to solve this so-called magnetic braking catastrophe
\citep{Joos2012,Li2013}, but not always.

This paper is a follow-up study of the transition region in the HH 111
protostellar system.  The properties of this system have been reported in
\citet{Lee2011} and only the important ones are summarized here.  This
system is deeply embedded in a compact molecular cloud core in the L1617
cloud of Orion at a distance of 400 pc.  At the center of this system, there
are two sources, VLA 1 and VLA 2, with a projected separation of $\sim$
\arcs{3} (1200 AU) and the former driving the prominent HH 111 jet
\citep{Reipurth1999}.  The VLA 1 source is a Class I protostellar system
with a flattened envelope, a rotating disk, and a highly collimated jet. 
Previous SMA observation of this system in \cCO{} has shown that the
flattened envelope is in transition to a KRD near the VLA 1 source
\citep{Lee2010}.  No clear influence of the VLA 2 source was seen on
the envelope and disk of the VLA 1 source. Here is a follow-up study of
this system with about 2 times higher angular and about 3 times higher
velocity resolutions in \cCO{} ($J=2-1$) obtained with ALMA at about 8 times
higher sensitivity.  In addition, to augment our study, we also include SO
($N_J=5_6-4_5$) shock emission at a similar angular resolution obtained with
the SMA.  This study shows for the first time the resolved region of the
transition, where a disk can be formed inside an envelope.


\section{Observations}


\subsection{ALMA Observations}

Observations toward the HH 111 system were obtained with ALMA using both the
12m array (in C32-4 configuration with a total time of $\sim$ 140 minutes)
and the 7m array (with a total time of $\sim$ 445 minutes).  This project
was a Cycle 1 project transferred to Cycle 2.  For the 12m array, 3
executions were carried out during 2014 April 13-28, all with 47.11 minutes
on the source.  33-35 antennas were used with  the projected baseline
length of 20$-$558.2 m. For the 7m array, 20 executions were carried out,
with 5 during 2013 December 14-15, 4 during 2014 January 11-15, 5 during
2014 April 09-May 03, and 6 during 2014 December 11-14, all with 22.37 min
on the target source, except for one with 19.47 min.  8-10 antennas were
used with the projected baseline length of 7$-$48.9 m. A 3-point
mosaic was used to cover $\sim$ \arcs{40} to the north and south from the
center, observing the envelope and disk in the equatorial plane of the
system.

The 230 GHz band receivers were used to observe the \aCO{} ($J=2-1$), \bCO{}
($J=2-1$), and \cCO{} ($J=2-1$) lines simultaneously with the 230 GHz
continuum.  Note that the \aCO{} and \bCO{} lines trace mainly the outflow
interaction and will be presented in a future publication.  The velocity
resolution in \cCO{} is $\sim$ 0.083 \vkm{}.

The data were calibrated with the CASA package, with Quasars J0750+1231 and
J0607-0834 as passband calibrators, Quasars J0532+0732 (a flux of 1.40$\pm
0.14$ Jy) and J0607-0834 (a flux of 0.63$\pm 0.07$ Jy) as gain calibrators,
and Callisto and Ganymede as flux calibrators.  With super-uniform
weighting and taper, the synthesized beam becomes circular
with a size of \arcsa{0}{50} in continuum. With super-uniform
weighting, the synthesized beam has a size of
\arcsa{0}{74}$\times$\arcsa{0}{62} at a position angle (P.A.)  of
77\degree{} in \cCO{}.  The rms noise level is $\sim$ 0.22 \mJyb{} ($\sim$
20 mK) for the continuum, and $\sim$ 4.3 \mJyb{} ($\sim$ 0.23 K) for the
\cCO{} channel maps at $\sim$ 0.083 \vkm{} velocity resolution.  
The velocities in the channel maps are LSR.  


\subsection{SMA Observations}

For SO observations with SMA, the details have been reported in
\citet{Lee2011} and are thus not repeated here.  The longest projected
baseline length is $\sim$ 480 m, similar to that of ALMA observations.  The
velocity resolution in SO line is $\sim$ 0.28 \vkm{} per channel.  Only one
single pointing toward the central region was observed to map the inner part
of the envelope and disk.  The visibility data were calibrated with the MIR
package.  The flux uncertainty was estimated to be $\sim$ 15\%.  The
calibrated visibility data was then imaged with the MIRIAD package, as
described in \citet{Lee2011}.  With super-uniform weighting, the synthesized
beam has a size \arcsa{0}{67}$\times$\arcsa{0}{57} with a P.A.  of
75\degree{}.  The rms noise levels are $\sim$ 35 \mJyb{} ($\sim$ 2.3 K) in
the channel maps.  The velocities in the channel maps are LSR.

\section{Results}




As in \citet{Lee2010}, the results are presented in comparison to a mosaic
image based on the Hubble Space Telescope (HST) NICMOS image ([FeII] 1.64
\micron{} + \H2{} at 2.12 \micron{} + continuum) obtained by
\citet{Reipurth1999}, which shows two VLA sources in the infrared,
reflection nebulae that trace the illuminated outflow cavity walls, and the
jet in the system.  The two sources have been detected at very high angular
resolution of $\sim$ \arcsa{0}{05} in 7 mm continuum by the VLA as the VLA 1
and 2 sources, respectively, at $\alpha_{(2000)}=\ra{05}{51}{46}{254}$,
$\delta_{(2000)}=\dec{+02}{48}{29}{65}$ and
$\alpha_{(2000)}=\ra{05}{51}{46}{07}$,
$\delta_{(2000)}=\dec{+02}{48}{30}{76}$ \citep{Rodriguez2008}.  These VLA
positions are adopted here as the source positions.   Based on the
fitting of the rotation curve in the envelope (Section \ref{sec:resC18O}),
the systemic velocity in this region is refined to be $\Vsys= 8.85\pm0.14$
\vkm{} LSR, from $8.9\pm0.14$ \vkm{} found in \citet{Lee2010} at 3 times
lower velocity resolution. Throughout this paper, we define an offset
velocity $\Voff = \VLSR - \Vsys$ to facilitate our presentation.


%

\subsection{230 GHz Continuum Emission} \label{ssec:cont}



\begin{figure} [!hbp]
\centering
\includegraphics[angle=-90,scale=1]{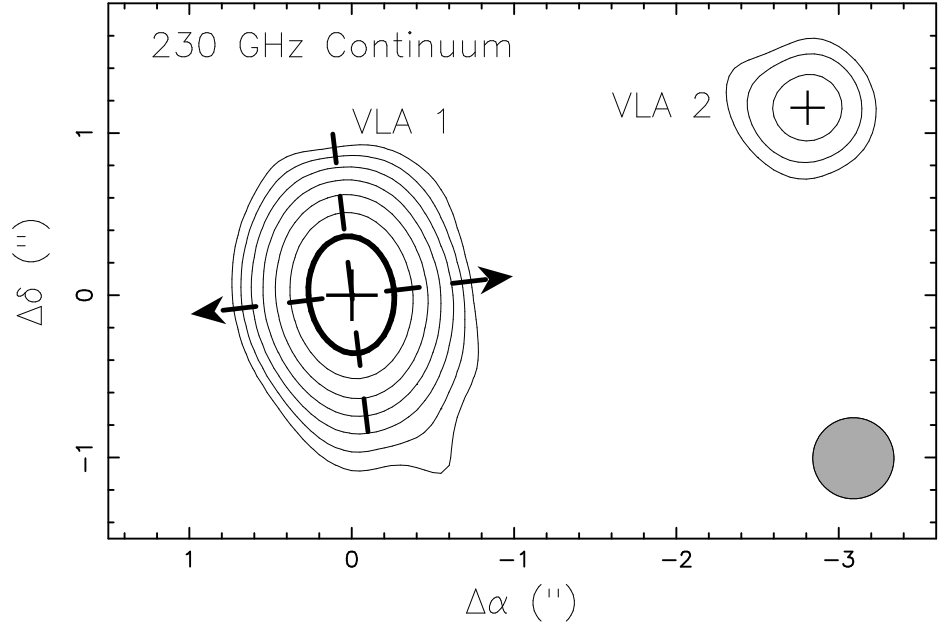}
\figcaption[]
{
The 230 GHz continuum map obtained with ALMA toward the VLA 1 and 2 sources.
The synthesized beam has a size of \arcsa{0}{5}.  The arrows indicate the
orientations of the blueshifted (western) and redshifted (eastern)
components of the jet, respectively, from the VLA 1 source.  
The dashed line indicates the
equatorial plane perpendicular to the jet axis.
The contour levels are $2^{(n-8)}P$,
where $P=173.8$ \mJyb{}, which is the peak value at VLA 1, and $n=1,2,..$.
The thick contour highlights the value at 50\% of the peak.
\label{fig:cont} } \end{figure}

Figure \ref{fig:cont} shows the 230 GHz continuum map toward the two VLA
sources at an angular resolution of \arcsa{0}{5}.  As discussed in
\citet{Lee2011}, the continuum emission is thermal dust emission, tracing
mainly the dusty disks around the two sources.  The continuum emission
toward VLA 1 is resolved into a disklike structure with a Gaussian
deconvolved size of \arcsa{0}{53}$\times$\arcsa{0}{19} (212 AU $\times$ 76
AU) and a P.A.  of $\sim$ 7\degree, with a flux density of 270$\pm$40 mJy,
all similar to those found in \citet{Lee2011}.  The orientation of this disk
is nearly exactly orthogonal to the jet axis, which has a P.A.  of
96.7\degree{}.  Interestingly, with unprecedented high ALMA sensitivity, we
can see clearly that the continuum emission intensity drops quickly
below 50\% of the peak beyond the Gaussian deconvolved radius ($\sim$
\arcsa{0}{27}, a half of the Gaussian deconvolved size along the major
axis). This suggests that the disk has a sharp outer boundary.  The
continuum emission toward VLA 2 is unresolved with a flux density of $\sim$
9$\pm$1.4 mJy.  As compared to \citet{Lee2011}, VLA 2 is better detected
because of higher sensitivity.



%

\subsection{\cCO{} Envelope and Disk}\label{sec:resC18O}

\begin{figure} [!hbp]
\centering
\includegraphics[angle=-90,scale=1]{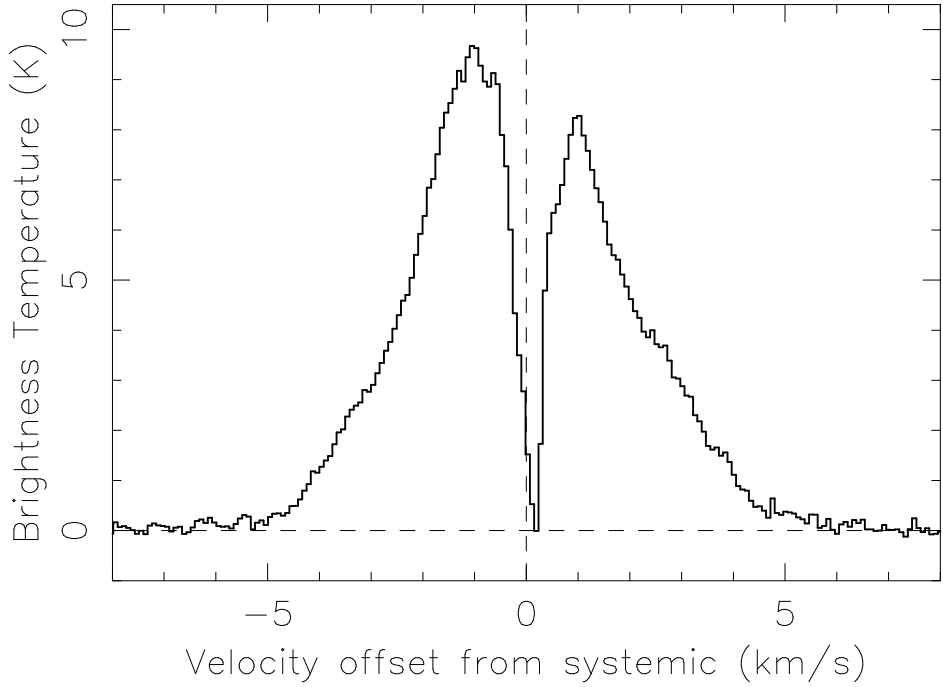} 
\figcaption[]
{\cCO{} spectrum toward the VLA 1 source averaged over an elliptical region of 
\arcs{2}$\times$\arcs{1} in size with the major axis in the equatorial plane.
The dashed vertical line marks the systemic velocity. The dashed horizontal line
marks the zero brightness temperature.
\label{fig:spec}
}
\end{figure}

Figure \ref{fig:spec} shows the \cCO{} spectrum toward the VLA 1 source
averaged over an elliptical region of \arcs{2}$\times$\arcs{1} elongated
along the equatorial plane (P.A.  $=$ 6.7\degree{}).  The spectrum shows a
double-peaked line profile with a blue asymmetry and an absorption dip
deepest at $\sim$  0.2 \vkm{}, similar to that seen before extracted
from \arcsa{1}{6} resolution SMA observation in \citet{Lee2010}.  The blue
asymmetry and absorption dip were used before to imply an infall motion in
the envelope \citep{Lee2010}.  The brightness temperature near the systemic
velocity is $\sim$ 25\% lower that found before,  likely because part
of the extended structure is resolved out in our observations at higher
angular resolution.  With ALMA sensitivity, the emission is now detected
more than $\pm$5 \vkm{} from the systemic, about 2 \vkm{} higher than that
observed before with the SMA.  As seen later, the emission at this high
velocity arises near the source, allowing us to better constrain the
Keplerian velocity near the source.

\begin{figure} [!hbp]
\centering
\includegraphics[angle=-90,scale=0.7]{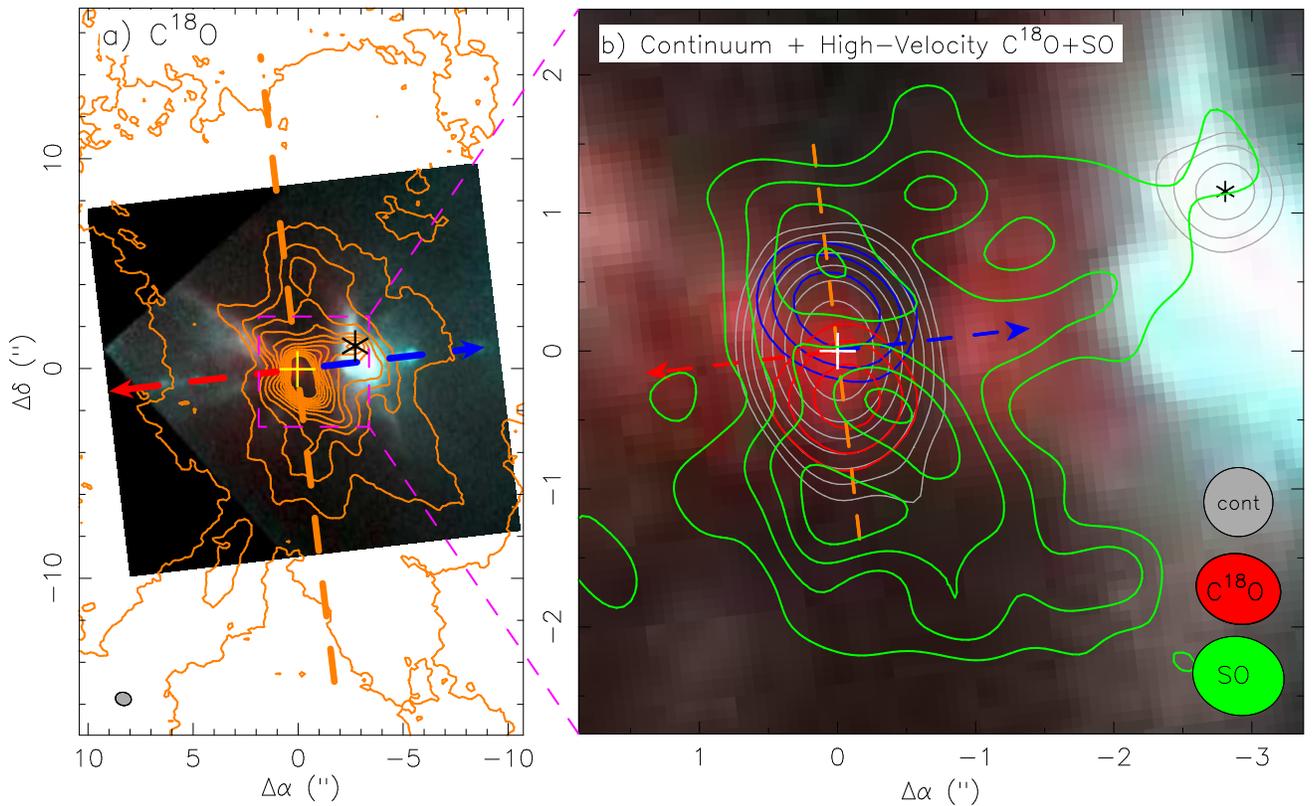} 
\figcaption[]
{\small
\tlabel{a}
Total intensity map of \cCO{} (orange contours)
on the HST NICMOS image in the HH 111 system.
Cross and asterisk mark the positions of the VLA 1 and 2 sources.
The contours start at 4$\sigma$ and have a step of 11$\sigma$, where
$\sigma=$ 0.0043 \Jybk{} 
\tlabel{b}
A zoom-in to the innermost part of the system around the VLA 1 source.
Gray contours show the dusty disk centered at the VLA 1 source, as 
in Figure \ref{fig:cont}.
Blue and red contours show the 
high-blueshifted ($\Voff =$ $-$3 to $-$6 \vkm{}) and 
high-redshifted ($\Voff= $ 3 to 6 \vkm{})
emission of \cCO{} on the disk. The contour levels are $3\cdot2^{(n-1)}\sigma$,
where $\sigma=2.2$ \mJyb{} \vkm{} and $n=1,2,..$.
Green contours shows the total intensity map of SO.
The contours start at 5$\sigma$ and have a step of 4$\sigma$, where
$\sigma=$ 0.04 \Jybk{}.
\label{fig:linemap}
}
\end{figure}

Figure \ref{fig:linemap}a shows the \cCO{} total intensity map on top of the
HST image.  The emission extends mainly along the equatorial plane of the
VLA 1 source, with some also extending to the NE, SE, SW, and NW around the
outflow cavity walls, and some also extending to the west to the VLA 2
source.  In the equatorial plane, \cCO{} emission shows an extended envelope
detected up to $\sim$ $\pm$\arcs{12} (4800 AU) from the VLA 1 source.  The
more extended emission detected in our previous SMA observations
\cite[extending to $\sim$ \arcs{16},][]{Lee2010} is resolved out from our
ALMA observations, due to a lack of shorter $uv$ coverage.  The emission
intensity increases toward the VLA 1 source, showing a flattened envelope
(with a thickness of $\sim$ \arcs{4}) formed within $\sim$ \arcs{6} of the
source in the extended envelope.  Interestingly, within this radius, the
rotation velocity was found to transition to Keplerian velocity
\citep{Lee2010}.  The intensity increases rapidly within $\sim$
\arcs{1}$-$\arcs{2} of the VLA 1 source, suggesting for a final change in
the structure from the (tenuous) flattened envelope to the (dense) disk. 
The envelope-disk is believed to be perpendicular to the jet that was
found to have an inclination of $\sim$ 10\degree{} to the plane of the sky
\citep{Reipurth1992}, and is thus almost edge-on, with its farside to the
west and nearside to the east.



\begin{figure} [!hbp]
\centering
\includegraphics[angle=0,scale=0.8]{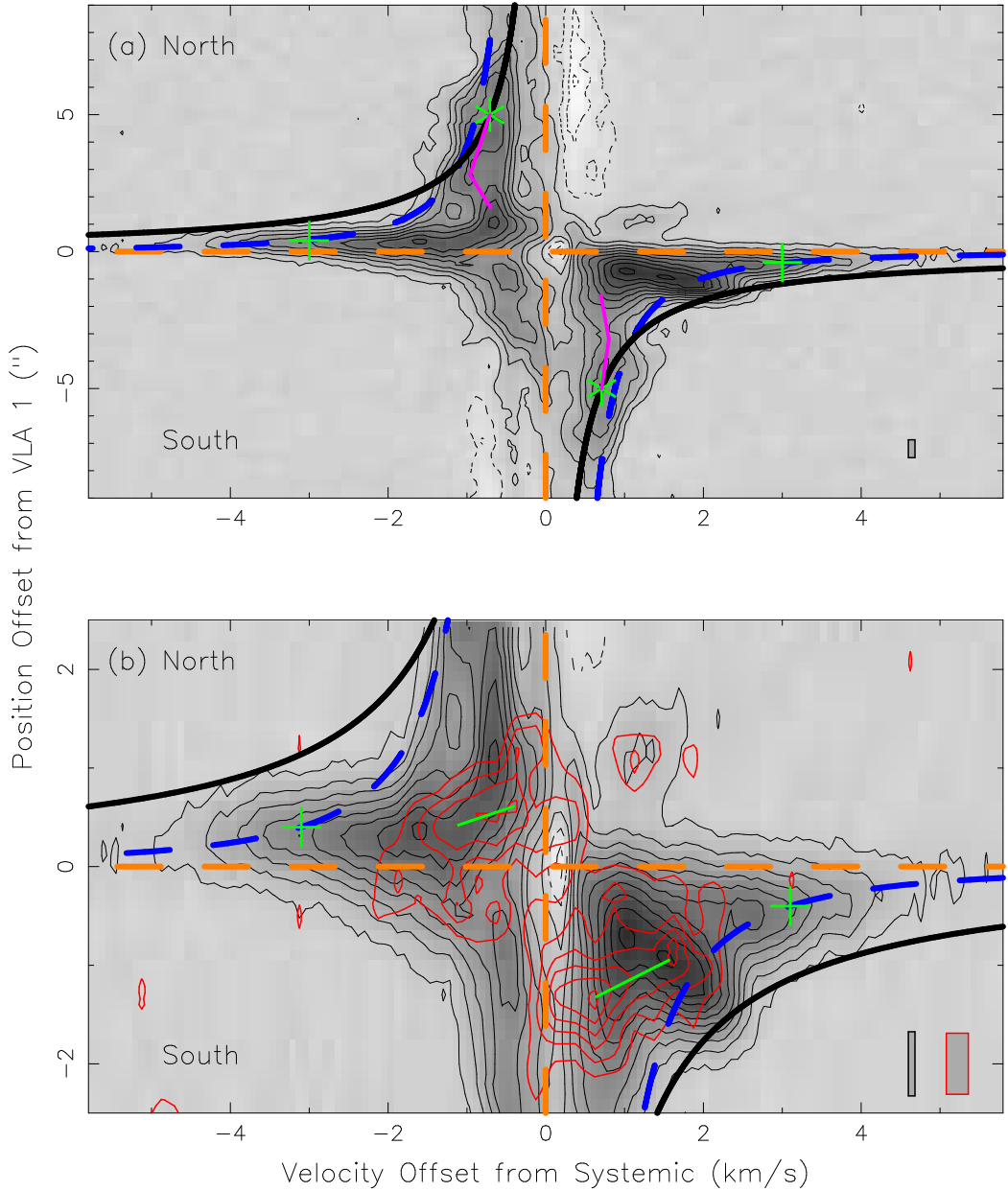} 
\figcaption[]
{\small
Position-Velocity (PV) diagrams in \cCO{} (gray image with black contours)
and SO [red contours in \tlabel{b}] centered at the VLA 1 source,
cut along the major axis of the envelope-disk.
For \cCO{}, the contours start at 3$\sigma$ with a step of 7$\sigma$,
where $\sigma=0.25$ K.
For SO, the contours start at 3$\sigma$ with a step of 2$\sigma$,
where $\sigma=2.2$ K.
The gray boxes in the lower-right corners show the velocity and
angular resolutions of the PV diagrams.
In (a), the magenta lines trace the change of rotation velocity within $\sim$ $\pm$\arcs{5} of the central source;
as we go toward the central source, the rotation velocity first increases with a rate slower than that in the outer part and
then decreases slightly in the region close to the disk.
Solid curves are derived from the rotation that has constant specific
angular momentum. Dashed curves are derived from the Keplerian rotation.
The asterisks mark roughly the inner radius where the rotation can be described 
by constant specific angular momentum. The crosses mark roughly the outer radius
where the rotation becomes Keplerian.
\label{fig:pv_env}
}
\end{figure}

Figure \ref{fig:pv_env} shows the position-velocity (PV) diagram of the
\cCO{} emission cut along the major axis of the envelope-disk in order to
study the rotation velocity in the envelope-disk.   In our previous
study with SMA, the rotation velocity was found to increase toward the
center, first with a profile corresponding to angular momentum conservation
in the outer part and then with a Keplerian velocity profile in the inner
part, with a change at the radius of $\sim$ \arcs{5} \citep{Lee2010}.  In
particular, the outer part ($r\sim$ \arcs{5}$-$\arcs{16}) could be fitted by
$v_\phi = v_c (\frac{r}{r_0})^{-1}$ with $v_c=3.9\pm0.4$ \vkm{}, while the
inner part ($r\lesssim$ \arcs{5}) by $v_\phi = v_k (\frac{r}{r_0})^{-0.5}$
with $v_k=1.75\pm0.2$ \vkm{}, where $r_0=$\arcs{1}.

Now we can refine the fitting parameters with ALMA at higher angular and
velocity resolutions.  The rotation velocity in the outer part is found to
follow the conservation of angular momentum down to $\sim$ $\pm$\arcs{5} as
found before (marked as asterisks), but with slightly smaller
$v_c=3.6\pm0.5$ \vkm{} (solid curves).  In this fit, the systemic velocity
is refined to 8.85 \vkm{} from 8.9 \vkm{}, in order to have a good fit on
both redshifted and blueshifted sides.  Note that the value of $v_c$ could
be smaller because the fit could be affected by the self-absorption and
missing flux near the systemic velocity.  However, the rotation velocity
within $\sim$ \arcs{5} of the central source does not change to Keplerian
immediately.  Instead, as we go toward the central source, the rotation
velocity first increases with a rate slower than that in the outer part and
then decreases slightly (as indicated by the magenta lines in Figure
\ref{fig:pv_env}a), and then increases rapidly toward the center, becoming
Keplerian within $\sim$ \arcsa{0}{4} of the central source.  The crosses at
(\arcsa{0}{4}, $-$3.1 \vkm{}) and ($-$\arcsa{0}{4}, 3.1 \vkm{}) mark roughly
the locations, within which the rotation velocity can be reasonably
described by the Keplerian velocity (dashed curves), with $v_k=2.0\pm0.3$
\vkm{}, slightly larger than that found before.  Therefore, unlike previous
study, only the very innermost part becomes a KRD.  The region in between
$\sim$ \arcs{5} and \arcsa{0}{4} can be considered as a transition region
between the envelope and the disk. The Keplerian velocity here implies a
mass of $\sim 1.8\pm0.5$ \solarmass{} for the central VLA 1 source.  Note
that there is a compact absorption dip centered at the VLA 1 source near the
systemic velocity, which gives the absorption dip in the spectrum shown in
Figure \ref{fig:spec}.   As discussed in \citet{Lee2014} for HH 212,
this compact absorption dip is likely due to an absorption of the continuum
emission of the disk by the near side of the envelope, which is found to be
infalling toward the center \citep{Lee2010}.

Figure \ref{fig:linemap}b shows the high-velocity emission of \cCO{},
greater than $\pm$3 \vkm{} from the systemic value, where the rotation
velocity becomes Keplerian, as discussed earlier.  The peaks of the
blueshifted and redshifted emission are on the opposites of the VLA 1 source
in the equatorial plane at a distance of $\sim$ \arcsa{0}{3}, coincident
with the radius of the dusty disk, further confirming that the disk is
a KRD.

\subsection{SO Shocks}

In order to further investigate the transition region, we also plot in
Figure \ref{fig:linemap}b the SMA map of SO emission, which was argued to
trace accretion shocks around the disk \citep{Lee2010}.  The SO emission was
detected within \arcs{2} of the VLA 1 source.  In the equatorial plane, two
SO emission peaks are seen on the opposite sides of the VLA 1 source, one at
$\sim$ \arcsa{0}{6} and one at $\sim$ $-$\arcsa{1}{2}, near the edge of the
high-velocity \cCO{} emission and dusty disk emission.  In the south, SO
emission is also seen closer in at $\sim$ $-$\arcsa{0}{4} (160 AU) extending
to the SW from the disk, and may trace material (i.e., slow wind) coming out
from the disk.  In addition, SO emission is also seen extending to the west,
probably tracing material connecting to the VLA 2 source.  SO emission is
also seen extending to the NW, probably tracing the envelope or outflow
around the cavity wall.


The PV diagram of the SO emission cut along the major axis is shown with the
red contours in Figure \ref{fig:pv_env}b.  The SO emission is detected
mainly within 2 \vkm{} of the systemic velocity.  The redshifted emission is
mainly in the south, and blueshifted emission is mainly in the north,
similar to those of the \cCO{} disk, suggesting that the shocked material
traced by the SO emission is also rotating.  On both blueshifted and
redshifted sides, the velocity of the emission increases toward the center
(as indicated by the green lines), connecting to the \cCO{} emission at
higher velocity, where the rotation velocity becomes Keplerian.  This
clearly suggests that SO shocks occur before the rotation velocity becomes
Keplerian and the disk is formed.  The SO emission closer in at
$-$\arcsa{0}{4} traces the base of the emission extending out from the disk
to the SW (see Figure \ref{fig:linemap}b).  It is seen on both blueshifted
and redshifted sides in the PV diagram, tracing a shock at the base in the
disk.




\subsection{Infall Motion} \label{ssec:infall}

\begin{figure} [!hbp]
\centering
\includegraphics[angle=-90,scale=0.7]{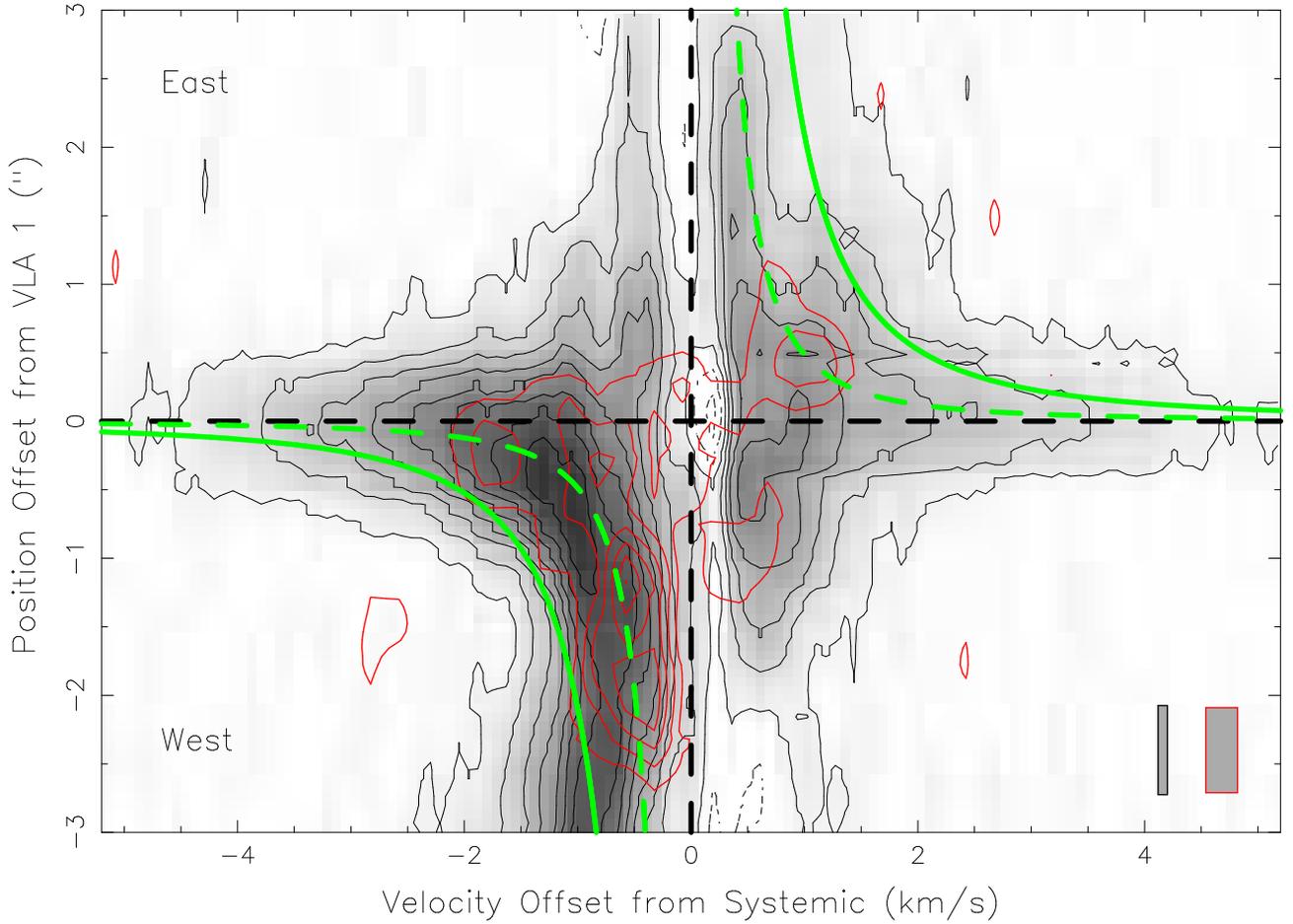} 
\figcaption[]
{PV diagrams in \cCO{} (black) and SO (red) centered at the VLA 1 source,
cut along the minor axis of the envelope-disk.
The contour levels are the same as those in Figure \ref{fig:pv_env}.
The gray boxes in the lower-right corners show the velocity and
angular resolutions of the PV diagrams.
Dashed curves show the infall velocity calculated with
$v_i = -1.7 (r/r_0)^{-0.5}$ \vkm{}, where $r_0$ is \arcs{1} (400 AU).
Solid curves show the free-fall velocity due to 
the central VLA 1 source.
\label{fig:pvinfall}
}
\end{figure}

Figure \ref{fig:pvinfall} shows the PV diagrams of \cCO{} and SO along the
minor axis of the envelope-disk,  in order to  refine the infall motion
in the envelope found in \citet{Lee2010}.  As mentioned earlier, the
envelope-disk is almost edge-on, with its farside to the west and nearside
to the east. The flattened envelope has a thickness of $\sim$ \arcs{4} and
a radius of $\sim$ \arcs{6}.  Thus, the \cCO{} emission beyond $\pm$
\arcs{3} from the center should be mainly from the outflow and is thus not
shown here.  In addition, the \cCO{} emission at $<-$\arcs{2} extends to the
west to the VLA 2 source and is thus affected by the source.  The SO
emission at $<-$\arcs{1} extends to the west to the VLA 2 source (see Figure
\ref{fig:linemap}b) and is thus also affected by the source.  As a result,
both of these emissions should be excluded when studying the infall motion
in the envelope.  In \cCO{}, the blueshifted emission of the envelope is
more on the west (farside) and the redshifted emission is more on the east
(nearside), consistent with a small infall motion toward the VLA 1 source. 
That the blueshifted emission is brighter than the redshifted emission also
supports the infall motion in the envelope.   Since SO shows a similar
velocity structure to \cCO{}, it can trace the infall motion as well.
Assuming that the infall velocity in the envelope increases toward the
center with $r^{-0.5}$ as in \citet{Lee2010}, then the infall velocity seen
in \cCO{} and SO can
be roughly fitted with $v_i =
-1.7 (r/r_0)^{-0.5}$ \vkm{} (dashed curves)\footnote{Note that it was thought to be $v_i=-0.7
(r/r_0)^{-0.5}$ \vkm{} in \citet{Lee2010}, due to wrong correction for
the projection effect.}. This infall velocity is $\sim$ 50\% of the
free-fall velocity due to the VLA 1 source (solid curves), which has a mass
of $\sim$ 1.8$\pm$0.5 \solarmass{} (as derived earlier from the Keplerian
velocity of the disk).






\section{Discussion}


\subsection{Flattened Envelope: Transition Region}

\begin{figure} [!hbp]
\centering
\includegraphics[angle=-90,scale=0.8]{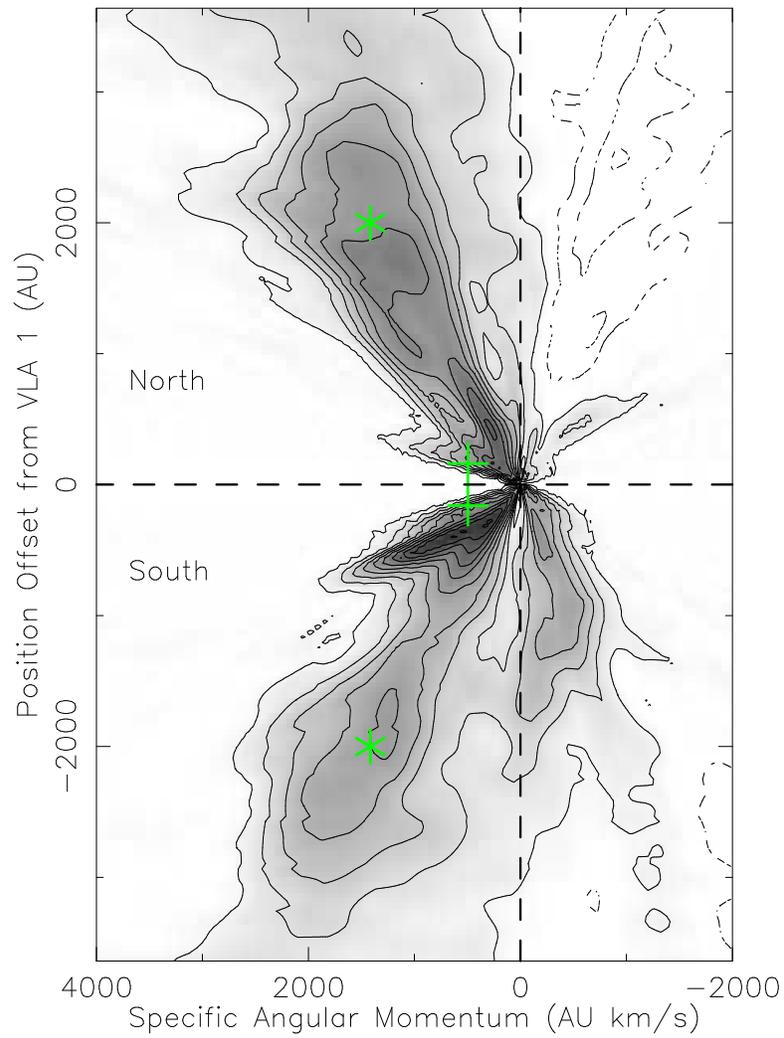} 
\figcaption[]
{Position-Angular momentum (PA) diagram in \cCO{} 
centered at the VLA 1 source, derived from Figure \ref{fig:pv_env}a. The specific angular momentum is derived
by multiplying the rotation velocity with the distance to the VLA 1 source.
The gray halftone and contour levels are the same as those in Figure \ref{fig:pv_env}.
The asterisks and crosses also have the same meanings as those in Figure \ref{fig:pv_env}.
\label{fig:PA}
}
\end{figure}

With ALMA observations in \cCO{} at higher angular and velocity resolutions,
we can better constrain how a KRD can be formed around a forming star.  The
\cCO{} map shows a flattened envelope embedded in a more extended envelope. 
It has a radius of $\sim$ \arcs{6} (i.e., 2400 AU), slightly larger than the
transition radius {($\sim$ \arcs{5} or 2000 AU)} where the rotation profile
starts to become flatter than that of conservation of angular momentum. 
However, the rotation velocity does not change immediately to that of
Keplerian at the transition radius.  Instead, the rotation velocity first
increases with a rate slower than that expected from conservation of angular
momentum, then decreases, and then increases rapidly near where the disk is
formed.  Therefore, the flattened envelope is mainly the transition region. 
The specific angular momentum, which is proportional to the product of
rotation velocity and radius, decreases tremendously in the transition
region,  by a factor of $\sim 3$, from a distance of roughly 2000 AU
(\arcs{5}) to 160 AU (\arcsa{0}{4}) (see Figure \ref{fig:PA} on the side of
positive angular momentum).  If the flow in the transition region
is in steady state, the drop implies a tremendous loss of specific angular
momentum, by a factor of 3. As a result, only a small KRD can be formed
near the central VLA 1 source.

In the flattened envelope, the infall velocity is not well determined,
because the flattened envelope is almost edge-on and thus not well resolved
along the minor axis.  Nonetheless, the infall velocity derived here should
be reasonable for the flattened envelope up to the transition radius, which
has a projected distance of $\sim$ \arcsa{0}{9}, resolvable with our
observations.  It is $\sim$ 50\% of the free-fall velocity due to the VLA 1
source.  Around the transition radius ($\sim$ \arcs{5}), it is $\sim$ 0.75
\vkm{}, roughly the same as the rotation velocity there, which is $\sim$
0.72 \vkm{}.  The complex rotation profile and substantial infall speed of
the flattened envelope indicate that it is the transition region between the
KRD (which is rotationally supported) and the more extended envelope, rather
than the KRD itself.

 The centrifugal radius of the envelope is where the centrifugal force
is balanced by the gravitational force and is thus where the rotation
velocity is equal to the Keplerian velocity.  In our model, this radius is
$(\frac{v_c}{v_k})^2 r_0=$\arcsa{3}{24} ($\sim$ 1300 AU), smaller than the
transition radius ($\sim$ \arcs{5}), but much larger than the disk radius. 
Hence, the gas of the envelope would never reach the disk radius unless its
angular momentum is reduced by some mechanisms.  We therefore believe that
the relatively large, $10^3$AU-scale, centrifugal radius of the envelope is
not directly associated with the formation of the much smaller rotationally
supported disk.  If it were the centrifugal radius directly responsible for
the KRD, the rotation speed inside the radius would increase toward the
center and the infall speed would quickly drop to zero if the total (infall
+ rotation) kinetic energy is conserved; these are inconsistent with our
observations.  The significant infall speed and apparent decrease in angular
momentum that we inferred in the transition region indicate the existence of
a second centrifugal radius closer to the central object that is more
directly responsible for the small KRD.  To study it in more detail, higher
resolution observations are needed.



\subsection{Keplerian Disk}


The radius of the disk can be roughly determined from our observations.  In
the innermost part of the flattened envelope, the rotation velocity becomes
Keplerian at a distance of $\sim$ \arcsa{0}{4} (e.g., 160 AU), forming a KRD
around the central source.  Since the PV structure is not well
resolved there, this distance can only be considered as a rough radius of the
disk.  The disk likely has a sharp outer boundary, with the continuum
emission decreasing rapidly beyond the disk radius (see Section
\ref{ssec:cont}).  On the other hand, in our previous observations at higher
angular resolution of $\sim$ \arcsa{0}{3} \citep{Lee2011}, the continuum
emission of the disk can be reproduced well using a flat disk model with a
radius of \arcsa{0}{6} (e.g., 240 AU).  Therefore, the radius of the disk
could reach up to 240 AU and higher angular resolution is really needed to
refine it.  In \citet{Lee2011}, the gas and dust associated with the disk
were estimated to have a total mass of $\sim 0.14 \pm 0.03$ \solarmass{}. 
Since this mass is only $\sim$ 8\% of the mass of the central VLA 1 source,
the disk should be gravitationally stable.


\subsection{A Ring of SO Shocks}

Interestingly, SO emission is detected in the innermost part of the
transition region around where the disk is formed, and where the rotation
velocity starts to increase again.  The SO emission appears as two peaks
outside the disk, one in the north at $\sim$ \arcsa{0}{6} (240 AU) and one
in the south at $\sim$ \arcsa{1}{2} (480 AU), in the equatorial plane.  The
SO material not only rotates (although with a lower velocity) in the same
sense as the disk, but also infalls toward the disk.  These two peaks are
likely to be two limb-brightened edges of a ring around the disk.  Since the
two peaks are at different distance from the VLA 1 source, the ring could be
elliptical.  As discussed in \citet{Lee2010}, SO abundance there is highly
enhanced as compared to that in the ambient medium.  Therefore, the SO
emission likely traces a ring of accretion shocks around the disk.

\subsection{Comparing to other sources}

Recent ALMA observations have also shown a transition of an infalling
envelope to a KRD even in Class 0 sources, e.g., L1527 IRS \citep{Sakai2014}
and HH 212 \citep{Lee2014}.  In L1527 IRS, a SO shock is also seen around
the disk at $\sim$ 100 AU.  Later study found that the rotation velocity
becomes Keplerian only further in at $\sim$ 40 AU \citep{Ohashi2014}.  This
supports the conclusion that the rotation velocity becomes Keplerian only at
the radius interior to the shocks, as found here in HH 111.  In HH 212, a
possible SO shock has been seen as well \citep{Podio2015}.  However, higher
velocity and angular resolutions are needed to check if there is also a
transition region with a decrease of rotation velocity and thus specific
angular momentum in these sources.



\subsection{Comparing to theoretical models}


In theoretical models of core collapse without magnetic field, e.g.,
\citet{Terebey1984} and \citet{Nakamura2000}, no such decrease of rotation
velocity would be seen before a KRD is formed.  However, it is well known
that both molecular clouds and their star-forming dense cores are magnetized
\cite[e.g., Planck Collaboration XIX, 2014;][]{Chapman2013,Hull2014}, and
realistic models should include magnetic fields, which can significantly
modify the dynamics of the core collapse and star formation, especially in
the transition region between the infalling envelope and the KRD that is
probed by our C$^{18}$O observations.

In the presence of a magnetic field of the strength typically measured in
dense cores \citep{Troland2008}, the envelope-disk transition region is
expected theoretically to be dynamically distinct from the rotationally
supported disk on the one hand, and the infalling envelope on the other. 
The expectation is based on the well-known magnetic flux problem in star
formation; namely, if all of the magnetic flux that threads a typical
star-forming core is carried into the formed star, the stellar field
strength would be orders of magnitude higher than the typically observed
value \cite[e.g.,][]{Shu1987}.  Only a tiny fraction of the core magnetic
flux can be dragged into the star; the vast majority of it must be decoupled
from the stellar material and be left behind.  \citet{Li1996} showed that,
in the idealized axisymmetric (2D) case, the decoupled magnetic flux is
trapped by the ram pressure of the infalling envelope behind an ambipolar
diffusion (AD) shock.  The trapped flux makes the field strength in the
post-AD-shock region much higher than in the pre-shock region.  The strongly
magnetized post-AD-shock region grows with time, reaching thousands of AUs
in size at late times, which is larger than the 100-AU scale KRD. 
Therefore, a generic expectation is that the KRD, where most of the
remaining (after magnetic braking) angular momentum originally associated
with the mass of the (single) star is stored, should be surrounded by a
strongly magnetized (envelope-disk) transition region, where most
of the magnetic flux originally associated with the same stellar mass is
parked.

The theoretically expected, strongly magnetized, envelope-disk transition
region provides a plausible explanation for the two most puzzling features
observed in the HH 111 system:  (1) the large decrease of the specific
angular momentum, by a factor of $\sim 3$, from a distance of roughly
2000 AU to 160 AU (see Figure \ref{fig:PA}), and (2) the slow infall
outside the KRD at a speed below the free-fall value (see Figure
\ref{fig:pvinfall}).
Both features can be naturally produced by a strong
magnetic field, which can remove angular momentum efficiently through
magnetic braking from field line twisting in the azimuthal direction, and
can slow down the gravitational collapse through magnetic tension force from
field pinching in the radial direction.  Both of these effects have been
illustrated through numerical simulations, most clearly under the assumption
of axisymmetry (2D).  For example, \citet{Li2011} found that the magnetic
field trapped in the post-AD-shock region could be so strong as to slow down
the infall speed temporarily to less than 10\% of the local free-fall value,
and remove essentially all of the angular momentum of the material passing
through the region (see their Figure 5).  However, exactly how much angular
momentum is removed by the magnetic braking and by how much the infall is
slowed down by the magnetic forces depend on many factors, such as the
degree of core magnetization and the level of ionization, which are
uncertain observationally for individual systems such as HH 111, and on
model simplifications.  For example, in 3D simulations (without the
axisymmetry assumption) of \citet{Krasnopolsky2012}, the strongly magnetized
post-AD-shock region becomes unstable to the ``magnetic interchange"
instability, which reduces the field strength somewhat compared to the 2D
case.  However, the field remains strong enough to slow down the infall
significantly and to remove angular momentum so efficiently as to prevent
the formation of a large, rotationally supported disk (see their Figure 5
for an illustration), at least at the relatively early times reached in
their simulations.  Additional physical processes, such as removal of small
grains \citep{Zhao2016} and reduction in cosmic ray ionization rate
\citep{Padovani2013}, and longer simulations are needed to produce
large 100 AU-scale disks around relatively evolved Class I sources that are
more suitable for direct comparison with the observations of the HH 111
system.  Nevertheless, its inferred sub-free-fall collapse and large loss of
angular momentum are qualitatively consistent with a strongly magnetized
envelope-disk region that is expected on general theoretical grounds.  This
interpretation can be tested by future high-resolution dust polarization and
Zeeman observations, perhaps with ALMA.

As for the SO shock, it  could be related to the accretion shock when the
infalling material right outside the KRD joins the KRD.  In this case, the
SO shock should be relatively thin.  Alternatively, it could be related to
heating through ambipolar diffusion in the post-AD-shock region, which could
be broader in the radial range.  The increase in rotation velocity from the
innermost part of the transition region to the outer part of the KRD could
be due to a redistribution of angular momentum on the disk, which tends to
transport mass toward the central star and angular momentum toward the outer
edge.

\section{Conclusions}



Our ALMA and SMA observations of the HH 111 protostellar system have shown
for the first time how a Keplerian rotating disk can be formed inside a
flattened envelope.  As in previous study, the flattened envelope is
detected in \cCO{}, extending out to $\gtrsim$ 2400 AU from the VLA 1
source.  It has a differential rotation, with the outer part ($\gtrsim$ 2000
AU) better described by a rotation that has constant specific angular
momentum and the innermost part ($\lesssim$ 160 AU) by a Keplerian rotation. 
The rotationally supported disk is therefore relatively compact in this
system, which is consistent with the dust continuum observations.  Most
interestingly, there is a substantial drop in specific angular momentum in
the envelope-disk transition region from 2000 AU to 160 AU, by a factor
of $\sim 3$, if the flow to the protostar is assumed to be in the steady
state. Such a decrease is not expected outside a disk formed from simple
hydrodynamic core collapse, but can happen naturally if the core is
significantly magnetized, because magnetic fields can be trapped in the
transition region outside the disk by the ram pressure of the protostellar
accretion flow, which can lead to efficient magnetic braking.  In addition,
SO shock emission is detected around the outer radius of the disk and could
trace an accretion shock around the disk.

\acknowledgements

We thank the anonymous referee for insightful comments.
This paper makes use of the following ALMA data:
ADS/JAO.ALMA\#2012.1.00013.S.  ALMA is a partnership of ESO (representing
its member states), NSF (USA) and NINS (Japan), together with NRC (Canada),
NSC and ASIAA (Taiwan), and KASI (Republic of Korea), in cooperation with
the Republic of Chile.  The Joint ALMA Observatory is operated by ESO,
AUI/NRAO and NAOJ.  C.-F.L.  and H.-C.H. acknowledge grants from the
Ministry of Science and Technology of Taiwan (MoST 104-2119-M-001-015-MY3)
and the Academia Sinica (Career Development Award).
ZYL is supported in part by NASA NNX14AB38G and NSF AST-1313083.


\clearpage

\end{document}